# Simulating Nuclear Dynamics with Quantum Effects

Sharon Hammes-Schiffer[1], Nancy Makri[2] and Mariana Rossi[3]

[1]Department of Chemistry, Yale University, New Haven, Connecticut 06520
[2]Department of Chemistry, Department of Physics, and Illinois Quantum Information Science and Technology Center, University of Illinois, Urbana, Illinois 61801
[3]Max Planck Institute for the Structure and Dynamics of Matter, Luruper Chaussee 149, 22761 Hamburg, Germany

**Status**

The structure and dynamics of molecules and materials, in all thermodynamic states, are determined by the laws of quantum mechanics. Solving various problems in this area requires a sufficiently accurate solution of the time-dependent or time-independent Schrödinger (or Dirac) equation for a system composed of many interacting electrons and nuclei. Under the much-celebrated Born-Oppenheimer (BO) approximation, the electronic problem has been addressed by a variety of approaches. Applications of these techniques have been central in the area of computational electronic structure theory.

Solving the equivalent equations for a system of interacting nuclei, as well as going beyond the BO approximation and obtaining the coupled dynamics of electrons and nuclei, is significantly more challenging. Accounting for nuclear quantum and non-BO effects can be far from just a small correction to a conventional calculation that considers nuclei as clamped point particles or as classical objects. Quantum nuclei have quantized energy levels, can tunnel through barriers, are delocalized, and can exhibit wave interference. Such quantum effects can dramatically change thermodynamic phase transitions, stabilize different crystal structures, influence the response of matter to stimuli, impact rates and equilibrium constants for chemical reactions, and cause isotope-dependent changes to the thermodynamics and kinetics. Thus, developing theoretical methods that incorporate quantum effects in nuclear dynamics is critical for answering many open questions in biology, physics, chemistry, and materials science.

A series of algorithmic developments, along with the increase of computer power has allowed quantum dynamics simulations of complex systems, triggering, in turn, the discovery of new situations where nuclear quantum dynamics are essential.[1] Available approaches can be broadly classified as those based on nuclear or nuclear-electronic wavefunctions, mixed quantum-classical approximations, and path integral (PI) methods. Each of these methods has advantages and limitations, as well as software implementations with varying degrees of accessibility. When choosing a method, one must balance accuracy and feasibility for the particular process of interest.

**Current and Future Challenges**

The ultimate goal of simulation methods is to treat all nuclei and all electrons quantum mechanically. For a wide range of important processes in chemistry and biology, this means accounting for zero-point energy (ZPE), nuclear tunneling, coherence, decoherence and quantum dissipation, treating the nuclear motion with full anharmonicity and accounting for changes in the electronic states (non-BO effects) when the nuclei rearrange.

An obvious difficulty in accounting for nuclear quantum effects in the dynamics of large molecular, biological, and condensed phase processes is the vast computational resources required to store and manipulate the quantum mechanical wavefunction. Finite-temperature effects pose an additional challenge to wavefunction-based methods when there are several low-frequency vibrational modes with many thermally populated states. Although a fully classical treatment of the nuclei cannot describe quantum effects such as hydrogen tunneling, in some cases treating (in addition to the coupled BO states) only one or a few nuclei (usually protons) by quantum mechanics is sufficient. The proper feedback among electronic states, quantum nuclei, and classical nuclei is important.

Treating the classical nuclei in terms of classical trajectories, which are local, while retaining a quantum treatment of electronic and/or some nuclear degrees of freedom, is possible through Ehrenfest's approximation, where the force on the classical particles is averaged with respect to the quantum wavefunction. Such a treatment can lead to unphysical results (for example, incorrect branching ratios). A significant improvement over Ehrenfest's approximation is achieved through surface hopping[2-3] (SH), which allows trajectories to hop between quantum states in a probabilistic fashion. Feynman's path integral (PI) formulation of quantum mechanics eliminates the need for delocalized wavefunctions, eliminating storage and allowing a consistent combination of quantum and classical treatments, but numerical integration of the resulting high-dimensional oscillatory function generally encounters serious convergence issues. When (as with normal mode vibrations, or through the validity of linear response) the nuclei can be treated as a harmonic bath coupled to the quantum system, the PI formulation offers a unique advantage, allowing a fully quantum mechanical treatment of all harmonic degrees of freedom, at zero or finite temperature, which can be evaluated using stable, numerically exact algorithms. The PI formulation in imaginary time offers an exact description of equilibrium processes with arbitrary potential functions, and efficient Monte Carlo and molecular dynamics methods are available for such calculations. This approach cannot describe time evolution but provides the basis for dynamical approximations.

**Advances in Science and Technology to Meet Challenges**

Fully quantum mechanical wavefunction propagation with many coupled degrees of freedom is often possible using the multiconfiguration time-dependent Hartree (MCTDH) methodology[4-5] This method converges to fully quantum mechanical results and has found many molecular applications. However, inclusion of a large number of relevant degrees of freedom and accounting for finite-temperature effects are generally not practical.

In hybrid approaches, specified nuclei are treated quantum mechanically, and the other nuclei are propagated on vibrational or vibronic surfaces with a nonadiabatic method such as SH. These approaches are useful for quantizing protons in simulations of proton transfer and proton-coupled electron transfer.[6] The nuclear-electronic orbital (NEO) approach treats specified nuclei, typically protons, quantum mechanically on the same level as the electrons with wave function or density functional theory methods.[7] The nuclear delocalization, ZPE, and tunneling of the quantum nuclei, as well as the anharmonic effects of the entire system, are inherently included. The nonadiabatic effects between the electrons and quantum nuclei are included without any BO separation, and the nonadiabatic effects of the classical nuclei with respect to the quantum subsystem can be included with Ehrenfest or SH dynamics. This approach enables real-time quantum dynamical simulations of thermal and photoexcited processes but neglects the quantum effects of the heavy nuclei.

For system-bath Hamiltonians, the quasi-adiabatic propagator path integral[8] (QuAPI) removes the instabilities arising from the oscillatory quantum phase, allowing numerically exact propagation. An analytically derived small matrix decomposition[9] (SMatPI) eliminates the QuAPI tensor storage, allowing calculations with many quantum states. The modular path integral[10] (MPI) extends these methods to large molecular aggregates, where each unit includes electronic states coupled to intramolecular vibrations. These real-time PI methods account for all interference and decoherence effects without approximation and have been used in many simulations of proton, electron and energy transfer.[11] The restriction to harmonic bath degrees of freedom is removed in the quantum-classical path integral[12] (QCPI), which captures the motion of the nuclei through classical trajectories that interact rigorously and consistently with the quantum subsystem.

The imaginary-time PI formalism for quantum statistical mechanics leads to useful and efficient (but mostly *ad hoc*) quantum dynamical approximations that can be applied to general anharmonic potentials, with a large number of quantum atoms at given thermodynamic conditions, and can be combined with electronic structure methods.[13] These methods combine quantum statistics with different types of classical time propagation, and can thus capture ZPE and incoherent tunneling effects, but completely miss quantum coherence. The recent development of Matsubara dynamics has exposed the relationship of centroid molecular dynamics[14] and (thermostatted[15]) ring polymer molecular dynamics[16] to quantum dynamics,[17] leading to new developments that improve these schemes. While it is straightforward to use these methods within the BO approximation and at equilibrium, there are many open challenges related to their extension to nonadiabatic and nonequilibrium situations.[18-19]

**Concluding Remarks**

This brief description of the challenges and advances for simulating quantum effects in nuclear dynamics cannot cover the rich history and diversity of this field, but instead focuses on a few successful approaches. Each approach has advantages and limitations, and methodological developments are underway to address the specific challenges. The goal of treating all nuclei and all electrons on equal footing beyond the BO approximation for realistic systems in a computationally practical way continues to be one of the most important frontiers in theoretical chemistry. Despite their limitations, however, the existing approaches enable simulations that provide useful insights into the physical mechanisms behind chemical and biological processes.

**Acknowledgements**

This material is based upon work supported by the National Science Foundation under Awards CHE-1954348 (to S.H.S.) and CHE-1955302 (to N.M.). M.R. acknowledges funding from the Max Planck Society and the Deutsche Forschungsgemeinschaft (DFG) - Projektnummer 182087777 - SFB 951.

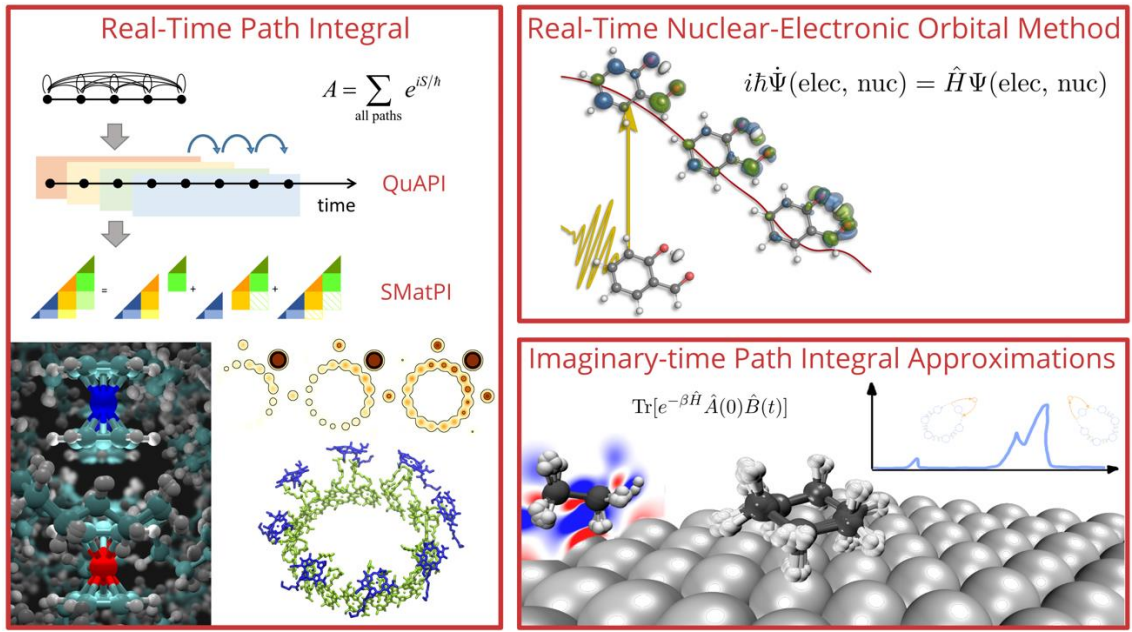

**Figure Caption**

**Left panel:** Schematic illustration of the path integral with an influence functional, the iterative QuAPI algorithm and the SMatPI decomposition. The image in the left bottom corner shows a snapshot of a QCPI simulation of electron transfer in the ferrocene-ferrocenium pair in liquid hexane, showing the solvent delocalization resulting from the superposition of three quantum-classical paths (adapted from Walters, P. L.; Makri, N., *J. Phys. Chem. Lett.* **2015,** *6*, 4959-4965). The yellow-brown contours in the right bottom corner of this panel are snapshots of the electronic density on the excited states of the 24 bacteriochlorophyll molecules in the B800-B850 LH2 complex of *Rhodopseudomonas molischianum* (with the two rings shown in blue and green), following excitation of a pigment on the B800 ring (adapted from Kundu, S.; Dani, R.; Makri, N., *Science Advances* **2022,** *8*, eadd0023). **Upper right panel:** Real-time NEO-TDDFT trajectory of excited state intramolecular proton transfer following photoexcitation to the $S_1$ electronic state. The time-dependent electron density difference relative to the ground state is shown as green (positive) and blue (negative) isosurfaces, and the time-dependent proton density is shown as a light gray isosurface. Details are given in Ref. Zhao, L.; Tao, Z.; Pavošević, F.; Wildman, A.; Hammes-Schiffer, S.; Li, X., *J. Phys. Chem. Lett.* **2020,** *11*, 4052-4058, and the figure was adapted from the associated journal cover. **Bottom right panel:** Snapshots of *ab initio* path-integral molecular dynamics simulations of cyclohexane on Rh(111), which captures electron-density rearrangements (blue and red regions). Details in Ref. K. Fidanyan, I. Hamada, and M. Rossi, *Adv Theory Simulations* **2021,** *4*, 2000241. Such simulations can be used to approximate real-time quantum correlation functions and calculate vibrational spectra, as sketched in the upper-right corner.